\documentclass[reprint,twocolumn,amsmath,amssymb,aps,prl,superscriptaddress]{revtex4-1}
\usepackage{graphicx}
\usepackage{dcolumn}
\usepackage{bm}
\usepackage{float}
\usepackage{amssymb}
\usepackage{color}
\usepackage{multirow}
\usepackage{stmaryrd}
\begin{document}
\title{Quantum oscillations in kagome metals (Ti, Zr, Hf)V$_6$Sn$_6$ at Van Hove filling}
\author{Miao He}
\affiliation{Anhui Key Laboratory of Low-Energy Quantum Materials and Devices, CAS Key Laboratory of Photovoltaic and Energy Conservation Materials, High Magnetic Field Laboratory of Chinese Academy of Sciences (CHMFL), HFIPS, CAS, Hefei 230031, China}
\affiliation{Science Island Branch of Graduate School, University of Science and Technology of China, Hefei 230026, China}

\author{Xitong Xu}
\email{xuxitong@hmfl.ac.cn}
\affiliation{Anhui Key Laboratory of Low-Energy Quantum Materials and Devices, CAS Key Laboratory of Photovoltaic and Energy Conservation Materials, High Magnetic Field Laboratory of Chinese Academy of Sciences (CHMFL), HFIPS, CAS, Hefei 230031, China}

\author{Ding Li}
\affiliation{Anhui Key Laboratory of Low-Energy Quantum Materials and Devices, CAS Key Laboratory of Photovoltaic and Energy Conservation Materials, High Magnetic Field Laboratory of Chinese Academy of Sciences (CHMFL), HFIPS, CAS, Hefei 230031, China}
\affiliation{Science Island Branch of Graduate School, University of Science and Technology of China, Hefei 230026, China}

\author{Qingqi Zeng}
\affiliation{Anhui Key Laboratory of Low-Energy Quantum Materials and Devices, CAS Key Laboratory of Photovoltaic and Energy Conservation Materials, High Magnetic Field Laboratory of Chinese Academy of Sciences (CHMFL), HFIPS, CAS, Hefei 230031, China}

\author{Yonglai Liu}
\affiliation{Anhui Key Laboratory of Low-Energy Quantum Materials and Devices, CAS Key Laboratory of Photovoltaic and Energy Conservation Materials, High Magnetic Field Laboratory of Chinese Academy of Sciences (CHMFL), HFIPS, CAS, Hefei 230031, China}
\affiliation{Science Island Branch of Graduate School, University of Science and Technology of China, Hefei 230026, China}

\author{Haitian Zhao}
\affiliation{Anhui Key Laboratory of Low-Energy Quantum Materials and Devices, CAS Key Laboratory of Photovoltaic and Energy Conservation Materials, High Magnetic Field Laboratory of Chinese Academy of Sciences (CHMFL), HFIPS, CAS, Hefei 230031, China}
\affiliation{Science Island Branch of Graduate School, University of Science and Technology of China, Hefei 230026, China}

\author{Shiming Zhou}
\affiliation{Hefei National Research Center for Physics Sciences at the Microscale, University of Science and Technology of China, Hefei 230026, China}

\author{Jianhui Zhou}
\affiliation{Anhui Key Laboratory of Low-Energy Quantum Materials and Devices, CAS Key Laboratory of Photovoltaic and Energy Conservation Materials, High Magnetic Field Laboratory of Chinese Academy of Sciences (CHMFL), HFIPS, CAS, Hefei 230031, China}

\author{Zhe Qu}
\email{zhequ@hmfl.ac.cn}
\affiliation{Anhui Key Laboratory of Low-Energy Quantum Materials and Devices, CAS Key Laboratory of Photovoltaic and Energy Conservation Materials, High Magnetic Field Laboratory of Chinese Academy of Sciences (CHMFL), HFIPS, CAS, Hefei 230031, China}
\affiliation{Science Island Branch of Graduate School, University of Science and Technology of China, Hefei 230026, China}

\begin{abstract}
Kagome materials have recently drawn great attention due to the interplay between nontrivial band topology, electron correlations, and Van Hove singularities related many-body orders.
Here we report three new vanadium-based kagome metals, TiV$_6$Sn$_6$, ZrV$_6$Sn$_6$, and HfV$_6$Sn$_6$, and conduct a comprehensive investigation of their structural, magnetic, and electrical transport properties.
All three compounds exhibit large unsaturated magnetoresistances and multiband Hall effects at low temperatures, indicating the existence of multiple highly mobile carriers.
Both the diagonal and off-diagonal resistivity show quantum oscillations with nontrivial Berry phases and high quantum mobilities.
First-principles calculations together with quantum oscillation analyses suggest the Van Hove singularities at the $M$ point for the three compounds all located in close vicinity of the Fermi level, and there also exist multiple topological nontrivial band crossings, including a nodal ring and a massive Dirac cone.
Our work extends the kagome \textit{AM$_6$X$_6$} family and paves the way for searching possible Van Hove physics in the V kagome lattice.
\end{abstract}

\date{\today}
\maketitle

\section{Introduction}
\begin{figure*}[htbp]
\begin{center}
\includegraphics[clip, width=0.85\textwidth]{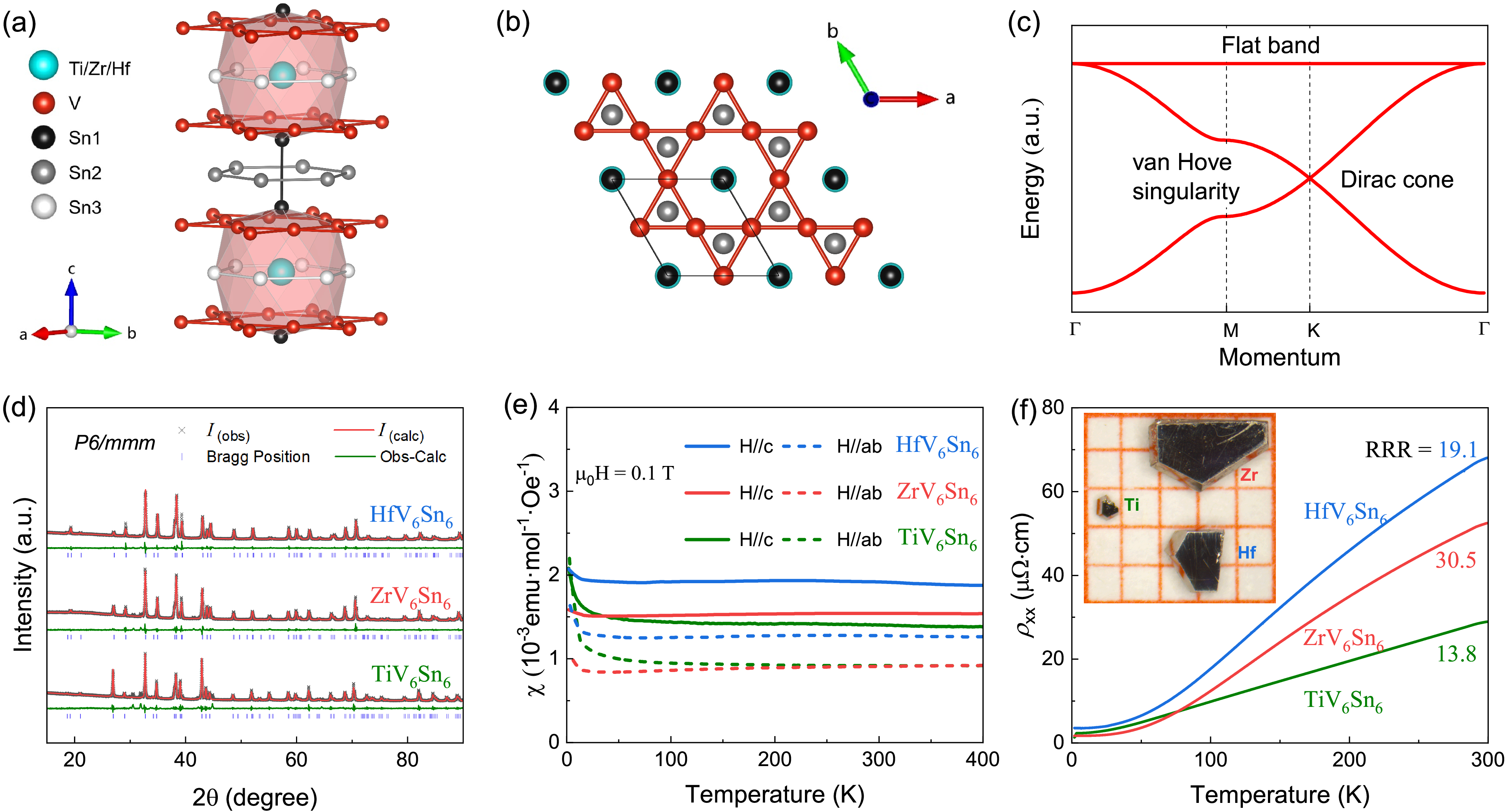}\\[1pt]
\caption{
(a) Crystal structure of (Ti, Zr, Hf)V$_6$Sn$_6$, which can be viewed as a CoSn-type framework caging the Ti, Zr, or Hf atoms (crystal structure created using VESTA~\cite{Momma2011}).
(b) Top view of the crystal structure, showing the kagome lattice of V and the triangular lattice formed by Ti, Zr, or Hf atoms.
(c) Tight-binding band structure of a kagome lattice showing Van Hove singularities at $M$ points, Dirac cones at $K$ points, and a flat band across the momentum space.
(d) Rietveld refinements for powder XRD pattern at room temperature.
(e) Magnetic susceptibility $\chi$ measured under 0.1~T along the $c$ axis and $ab$ plane.
(f) Temperature dependence of electrical resistivity $\rho_{xx}$ at zero magnetic field.
Inset: An optical photo of single crystals with millimeter-size hexagonal (001) facets for (Ti, Zr, Hf)V$_6$Sn$_6$.
}
\label{f1}
\end{center}
\end{figure*}

The kagome lattice, a two-dimensional network made of corner-sharing triangles, has been recognized as an optimal toy model for investigating emergent quantum phenomena~\cite{sy1951,Simeng2011,Yin2021,Yin2022Topological,Xu2023}.
Owning to its special symmetry, the electronic band inherently hosts Dirac crossing, Van Hove singularities (vHSs), and flat bands as shown in Fig.~\ref{f1}(c), giving birth to both nontrivial band topology and electron correlation~\cite{Ye2018Massive,Liu2020,Kang2020Dirac,Yin2020,Teng2023,wilson2023av3sb5}.
Of particular interest is the Van Hove singularity-related many-body orders~\cite{Ortiz2020,Jiang2021Unconventional,Kang2022,Rina2022}.
Previous theoretical studies predicted that at the Van Hove filling the kagome lattice would host competing electronic orders and Fermi surface instabilities~\cite{Wang2013,Kiesel2013}.
These phenomena have been demonstrated in recent \textit{A}V$_3$Sb$_5$ (\textit{A} = K, Cs, Rb) where charge density waves (CDW) and unconventional superconductivity coexist~\cite{Jiang2021Unconventional,Kang2022,Lin2021}.
Furthermore, the \textit{A}V$_3$Sb$_5$ family has also been shown to exhibit other exotic phases, including the nematic order~\cite{Nie2022}, the chiral charge order~\cite{Jiang2021Unconventional}, and pair density wave~\cite{Chen2021}.

Compared with the relatively small \textit{A}V$_3$Sb$_5$ family, the intermetallic \textit{AM$_6$X$_6$} (\textit{A} = alkali, alkali earth and rare earth metals, \textit{M} = transition metals; and \textit{X} = Sn, Ge, etc.) offers more tunability~\cite{Xu2023,wilson2023av3sb5}.
This family, containing over 100 compounds, features an \textit{M}-based kagome lattice and additional diversity in the \textit{A} and \textit{X} sites~\cite{Venturini2006,fredrickson2008origins,Baranov2011}.
In the \textit{M} = Mn- or Fe-based \textit{AM$_6$X$_6$} systems, the kagome lattice can exhibit novel magnetism, magnetic topological phases, and strong correlation, with TbMn$_6$Sn$_6$ being representative~\cite{Yin2022Topological,Xu2022,Ma2021rare,Ma2021anomalous,Lee2023interplay}.
In the V-based nonmagnetic \textit{R}V$_6$Sn$_6$ system (\textit{R} = rare earth metals), the basic feature of a kagome lattice is retained and enriched by possible local magnetism from the \textit{R} site.
For instance, topological Dirac surface states and vHSs have been observed by angle-resolved
photoemission spectroscopy (ARPES) experiments in GdV$_6$Sn$_6$~\cite{Peng2021,Pokharel2021,Yong2022}.
Strong uniaxial ferromagnetism and the corresponding anomalous Hall effect has been unveiled in TbV$_6$Sn$_6$~\cite{Rosenberg2022}.
In YbV$_6$Sn$_6$, a quantum critical behavior based on a Yb-triangular Kondo lattice has been also demonstrated~\cite{Guo2023}.
Specifically, ScV$_6$Sn$_6$ is the unique material within the \textit{AM$_6$X$_6$} system that has been found to show a CDW phase up to now~\cite{Arachchige2022}.
The relation between the observed ($\frac{1}{3},\frac{1}{3},\frac{1}{3}$) CDW and the vHSs in the electronic band has garnered significant attention recently~\cite{hu2023phonon,lee2023nature}.

In this work we have synthesized three new V-based kagome compounds, TiV$_6$Sn$_6$, ZrV$_6$Sn$_6$, and HfV$_6$Sn$_6$.
The single crystals, with a Pauli paramagnetic ground state, all reveal a large unsaturated magnetoresistance and multiband Hall effect at low temperatures.
Large Shubnikov-de Haas quantum oscillations (SdH-QOs) with nontrivial Berry phases have been observed in all three compounds.
The near-zero effective mass (0.035~$m_e$) and exceptionally high quantum mobility ($\sim5\times10^3~\mathrm{cm^2/Vs}$) in ZrV$_6$Sn$_6$ and HfV$_6$Sn$_6$ strongly indicate a relativistic origin.
Remarkably, the band-structure calculations and quantum oscillation parameters suggest that the vHS at the $M$ point for Zr and Hf siblings locates very close to the Fermi level (20 meV for Ti and less than 5 meV for Zr and Hf).
Such a proximity is thought to be responsible for the many-body electronic orders in the \textit{A}V$_3$Sb$_5$ family~\cite{Kang2022,Rina2022}.
Although no signs of the electronic instabilities are observed in the as-grown (Ti, Zr, Hf)V$_6$Sn$_6$ crystals, it is possible that the vHS-related physics could be revealed by a fine tuning of the Fermi level. Our findings therefore provide an opportunity for further pursuing the vHS-related physics in the V-based kagome metals.

\section{Experimental techniques}
Single crystals of (Ti, Zr, Hf)V$_6$Sn$_6$ were synthesized via a Sn self-flux method.
Prior to the growth, Ti/Zr/Hf and V grains were arc-melted together in order to achieve better homogeneity.
The materials, mixed in a molar ratio of 1:6:50, were loaded inside an alumina crucible and sealed in an argon-filled fused silica ampoule.
The mixtures were maintained at 1125$^{\circ}$C over a 24-h period, after which they were slowly cooled to 780$^{\circ}$C at a rate of 2$^{\circ}$C/h.
Shiny, hexagonal-shaped single crystals were isolated from the tin-flux through centrifugation [inset of the Fig.~\ref{f1}(f)].

The crystal structures of (Ti, Zr, Hf)V$_6$Sn$_6$ were determined through both single-crystal and powder x-ray diffractions (XRD) measurements in a SuperNova Rigaku single-crystal diffractometer and a Rigaku MiniFlex instrument, respectively.
Chemical composition was further ascertained through energy-dispersive x-ray spectroscopy (EDX) measurements, yielding Ti$_{1.16}$V$_{5.80}$Sn$_6$, Zr$_{1.06}$V$_{5.94}$Sn$_6$, and Hf$_{0.95}$V$_{6.09}$Sn$_6$, respectively, which closely aligns with the expected 166 stoichiometry (Fig.~S2) (see Supplemental Material SM~\cite{SM}).
For the Ti compound, there is evidence for slight off-stoichiometry between the Ti and V ratios in EDX.
It is challenging to distinguish between the sites using x-ray techniques, and there is potential for Ti-V disorder within crystals of TiV$_6$Sn$_6$.
However, for brevity we refer to this compound using the stoichiometric nomenclature for the remainder of the manuscript.
Magnetization and electrical transports were measured in a Quantum Design MPMS3-7~T and a Cryomagnetics C-MAG 12~T system, respectively.

\begin{table}[htbp]
\footnotesize
\caption{\label{t1}
Lattice parameters for (Ti, Zr, Hf)V$_6$Sn$_6$ at room temperature.
}
\begin{tabular}{p{2cm}<{\centering}p{2cm}<{\centering}p{2cm}<{\centering}p{2cm}<{\centering}}
\hline\hline
                       &   TiV$_6$Sn$_6$          &   ZrV$_6$Sn$_6$          &   HfV$_6$Sn$_6$ \\
\hline
$a$~({\AA})            &   5.4452 (2)             &   5.4602 (2)             &   5.4565 (5)  \\
$c$~({\AA})            &   9.1916 (4)             &   9.1634 (4)             &   9.1678 (11) \\
Volume~({\AA}$^3$)     &   236.02 (2)             &   236.59 (2)             &   236.39 (6)  \\
\hline\hline
\end{tabular}
\end{table}

DFT calculations with (w/) and without (w/t) spin-orbit coupling (SOC) were performed with the Perdew-Burke-Ernzerhof (PBE) exchange-correlation functional using a plane-wave basis set and projector augmented wave method, as implemented in the Vienna Ab initio Simulation Package (VASP)~\cite{Perdew1996,P.E.1994,Kresse1996}.
A plane-wave basis set with a kinetic energy cutoff of 350~eV is considered while performing first-principles calculations.
A $\Gamma$-centered Monkhorst-Pack ($9\times9\times5$) $k$-point mesh was adopted for the Brillouin zone sampling and smearing of 0.1~eV.

\section{Results and discussions}
\begin{figure}[h]
\centering
\includegraphics[clip, width=0.47\textwidth]{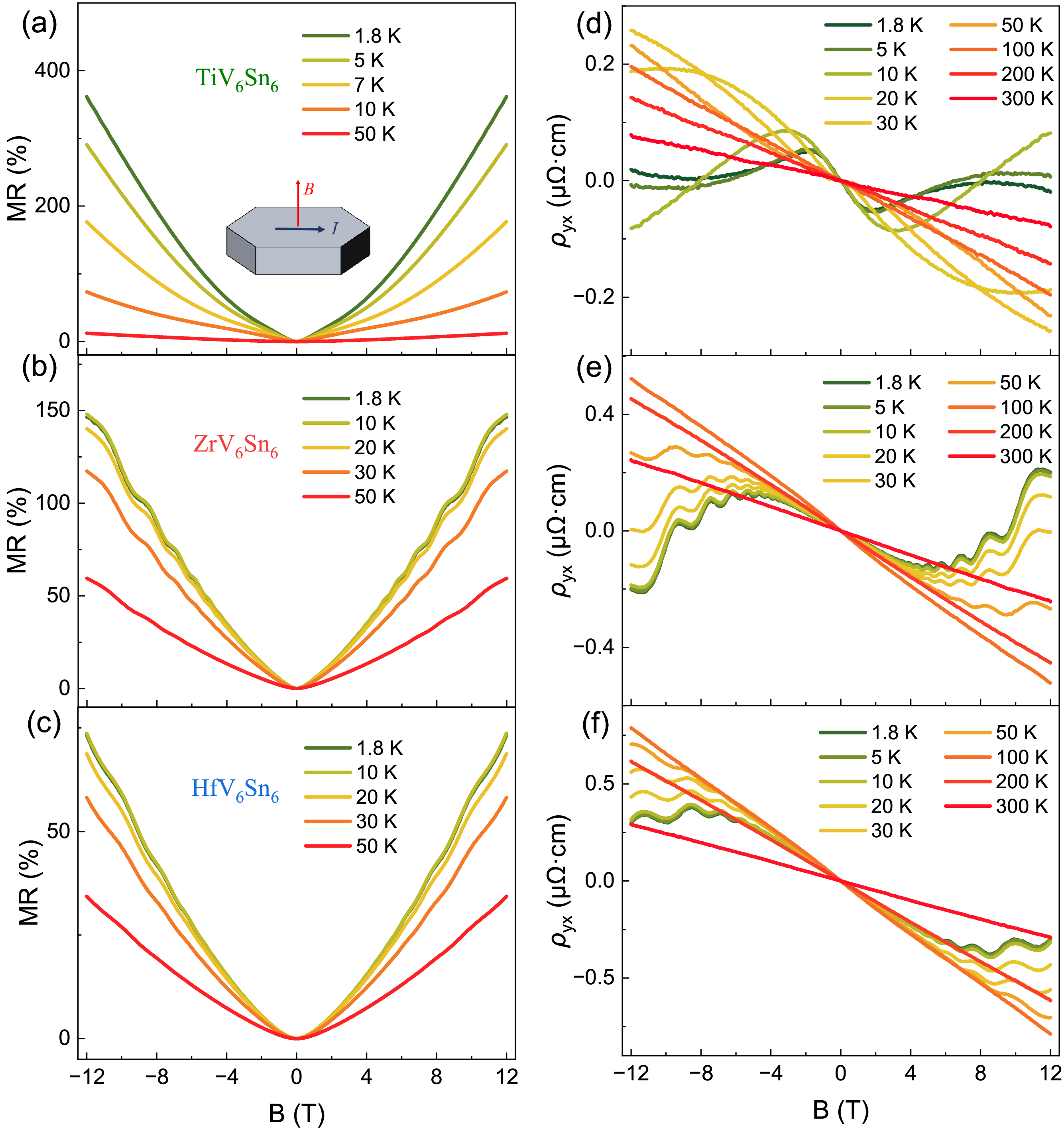}
\caption{
Field dependence of (a)-(c) MR, and (d)-(f) Hall resistivity $\rho_{yx}$ at various temperatures for (Ti, Zr, Hf)V$_6$Sn$_6$.
The magnetic field is applied perpendicular to the $ab$ plane, as shown in the inset of (a).
}
\label{f2}
\end{figure}

Similar to the known \textit{A}M$_6$Sn$_6$ compounds, TiV$_6$Sn$_6$, ZrV$_6$Sn$_6$, and HfV$_6$Sn$_6$ crystallize in the hexagonal HfFe$_6$Ge$_6$-type structure (space group $P6/mmm$, No.~191).
This framework can be regarded as a stuffed version of the CoSn structure as shown in Fig.~\ref{f1}(a)~\cite{Venturini2006,fredrickson2008origins}.
An alternation of Sn-centered V kagome nets with Sn honeycomb nets creates large hexagonal void spaces in the cage, serving as host to the Ti/Zr/Hf ions.
As the chemical pressure from the cations pushes the Sn sites within the kagome nets away from the void center, this structure highlights a pristine V-based kagome within the $ab$ plane [Fig.~\ref{f1}(b)], which plays a pivotal role in constructing the kagome physics observed in analog \textit{R}V$_6$Sn$_6$~\cite{Peng2021,Pokharel2021,Yong2022,Guo2023,Arachchige2022,meier2023tiny}.
The refinements of powder XRD patterns are shown in Fig.~\ref{f1}(d) and Fig.~S1~\cite{SM}, showing good consistency between experiments and calculations.
The detailed single-crystal lattice parameters from single-crystal XRD refinements are summarized in Table~\ref{t1} and SM Table~S1~\cite{SM}.

The temperature-dependent magnetic susceptibility under an external field of 0.1~T along the $c$ axis and $ab$ plane for (Ti, Zr, Hf)V$_6$Sn$_6$ is shown in Fig.~\ref{f1}(e).
The overall profiles of all curves resemble each other.
The three compounds exhibit predominantly Pauli-paramagnetic behavior up to 400~K and a small Curie tail at low temperatures, which is similar to that of their nonmagnetic sibling YV$_6$Sn$_6$~\cite{Pokharel2021,Zhang2022PRM}.
Figure~\ref{f1}(f) depicts the temperature dependence of the resistivity $\rho_{xx}$, measured with current applied within the $ab$ plane.
The $\rho_{xx}$ at room temperature increases from Ti to Hf.
All the compounds exhibit typical metallic behavior with residual resistance ratios [RRRs, defined as $\rho_{xx}$ (300~K)/$\rho_{xx}$ (2~K)] ranging from 14 to 30 for Ti, Hf, and ZrV$_6$Sn$_6$ in the \textit{R}V$_6$Sn$_6$ system.
These values are much larger than the nonmagnetic counterparts in the \textit{R}V$_6$Sn$_6$ family~\cite{Arachchige2022,Guo2023,Pokharel2021}, indicating the high quality of our samples.

Figure~\ref{f2} shows the magnetoresistance (MR, [$\rho_{xx} (B)$-$\rho_{xx} (0)$]/$\rho_{xx} (0)$$\times$100\%) and the Hall resistivity $\rho_{yx}$ at representative temperatures, with magnetic field applied along the crystallographic $c$ direction.
At base temperature, the MR for the three compounds is unsaturated at 12~T, ranging from 70\% to 360\% in Hf, Zr, and TiV$_6$Sn$_6$.
The $\rho_{yx}$ for the three compounds shares a similar trend.
At temperatures above 100~K, all of the $\rho_{yx}$ exhibit a negative, nearly field-linear behavior, indicating the majority charge carriers are the electrons.
However, a nonlinear behavior emerges at low temperatures, and there is a steplike feature for TiV$_6$Sn$_6$.
This is reminiscent of the possible anomalous Hall response in the \textit{A}V$_3$Sb$_5$~\cite{doi:10.1126/sciadv.abb6003,PhysRevB.104.L041103}.
However, when we look at the off-diagonal conductivity $\sigma_{xy}$ ($\sigma_{xy}=\frac{\rho_{yx}}{(\rho_{xx}^2+\rho_{yx}^2)}$), the overall profile can be well fitted using a two-band model~\cite{hurd2012},
$$\sigma_{xy}=\left\lbrack n_h\mu_h^2 \frac{1}{1+(\mu_hB)^2}-n_e\mu_e^2 \frac{1}{1+(\mu_eB)^2}\right\rbrack eB,$$
where $n_h$($n_e$) and $\mu_h$($\mu_e$) denote the carrier concentration and mobility of holes (electrons), respectively.
By fitting the two-band model, we subtracted carrier mobilities all about $10^3 \mathrm{cm^2/Vs}$, as shown in Fig.~S5~\cite{SM}.
We therefore attribute the temperature evolution of the Hall responses to the competition between highly mobile electrons and holes.

\begin{figure*}[htbp]
\begin{center}
\includegraphics[clip, width=0.85\textwidth]{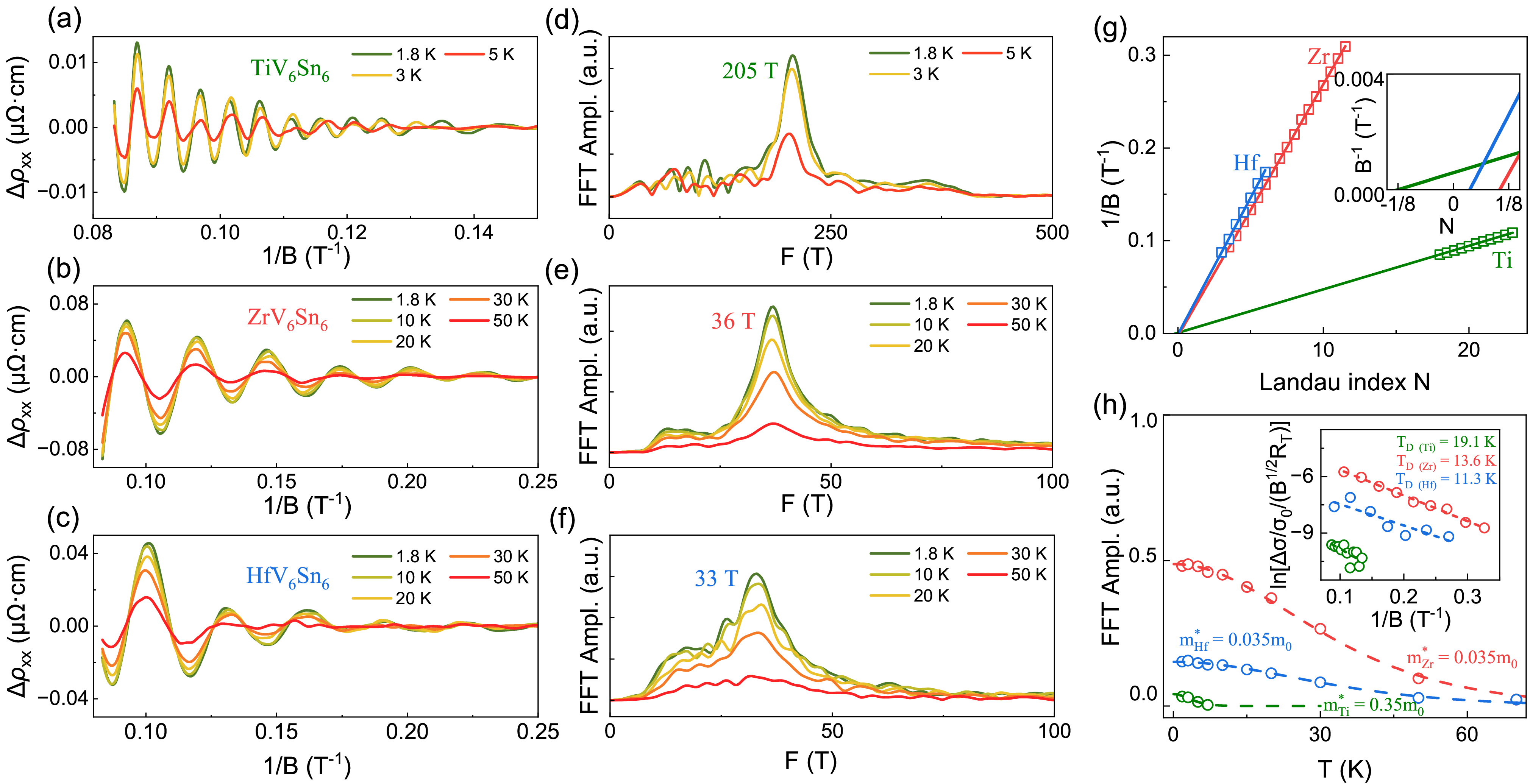}\\[1pt]
\caption{
(a)-(c) Quantum oscillations in $\rho_{xx}$ after subtracting a smooth background for (Ti, Zr, Hf)V$_6$Sn$_6$.
(d)-(f) Corresponding FFT spectra of SdH oscillations.
(g) The Landau fan diagram derived from the SdH oscillations at 1.8~K.
Inset is a zoom-in near $N$ = 0.
(h) Temperature dependences of FFT peak amplitudes.
The cyclotron masses were extracted from the fitted curves using the LK formula.
Inset shows the Dingle plots of SdH oscillation amplitudes at 1.8~K.
}
\label{f3}
\end{center}
\end{figure*}

\begin{table*}[htbp]
\footnotesize
\caption{\label{t2}
Quantum oscillation parameters for (Ti, Zr, Hf)V$_6$Sn$_6$. $F$, the oscillation frequency; $m^*$, cyclotron mass; $A_F$, Fermi cross section area; $k_F$, Fermi vector; $\nu_F$, Fermi velocity; $T_D$, Dingle temperature; $\tau_q $, quantum relaxation time and $\mu_q$, the quantum mobility.
}
\begin{tabular}{p{1.5cm}<{\centering}p{1.5cm}<{\centering}p{2cm}<{\centering}p{2cm}<{\centering}p{2cm}<{\centering}p{2cm}<{\centering}p{2cm}<{\centering}p{2cm}<{\centering}p{2cm}<{\centering}}
\hline\hline
                  &   $F$ (T)        &   $m^*$ ($m_0$)       &   $A_F$ (${\AA}^{-2}$)      &   $k_F$ (${\AA}^{-1}$)      &   $\nu_F$ (10$^5$~m/s)  &   $T_D$ (K)     &   $\tau_q$ (10$^{-14}$~s)     &   $\mu_q$ (cm$^2$/Vs)\\
\hline
TiV$_6$Sn$_6$     &   205$\pm$2      &   0.35$\pm$0.01       &   0.020$\pm$0.001           &   0.079$\pm$0.001           &   2.61$\pm$0.01         &   19.1$\pm$2.5  &   6$\pm$1                     &316$\pm$30\\
ZrV$_6$Sn$_6$     &   36$\pm$1       &   0.035$\pm$0.001     &   0.003$\pm$0.001           &   0.033$\pm$0.001           &   10.9$\pm$0.1         &   13.6$\pm$0.5  &   8.8$\pm$0.1                 &4433$\pm$150\\
HfV$_6$Sn$_6$     &   33$\pm$1       &   0.035$\pm$0.001     &   0.003$\pm$0.001           &   0.032$\pm$0.001           &   10.5$\pm$0.1          &   11.3$\pm$1.5  &   10$\pm$1                    &5335$\pm$500\\
\hline\hline
\end{tabular}
\end{table*}

\begin{figure}[h]
\centering
\includegraphics[clip, width=0.4\textwidth]{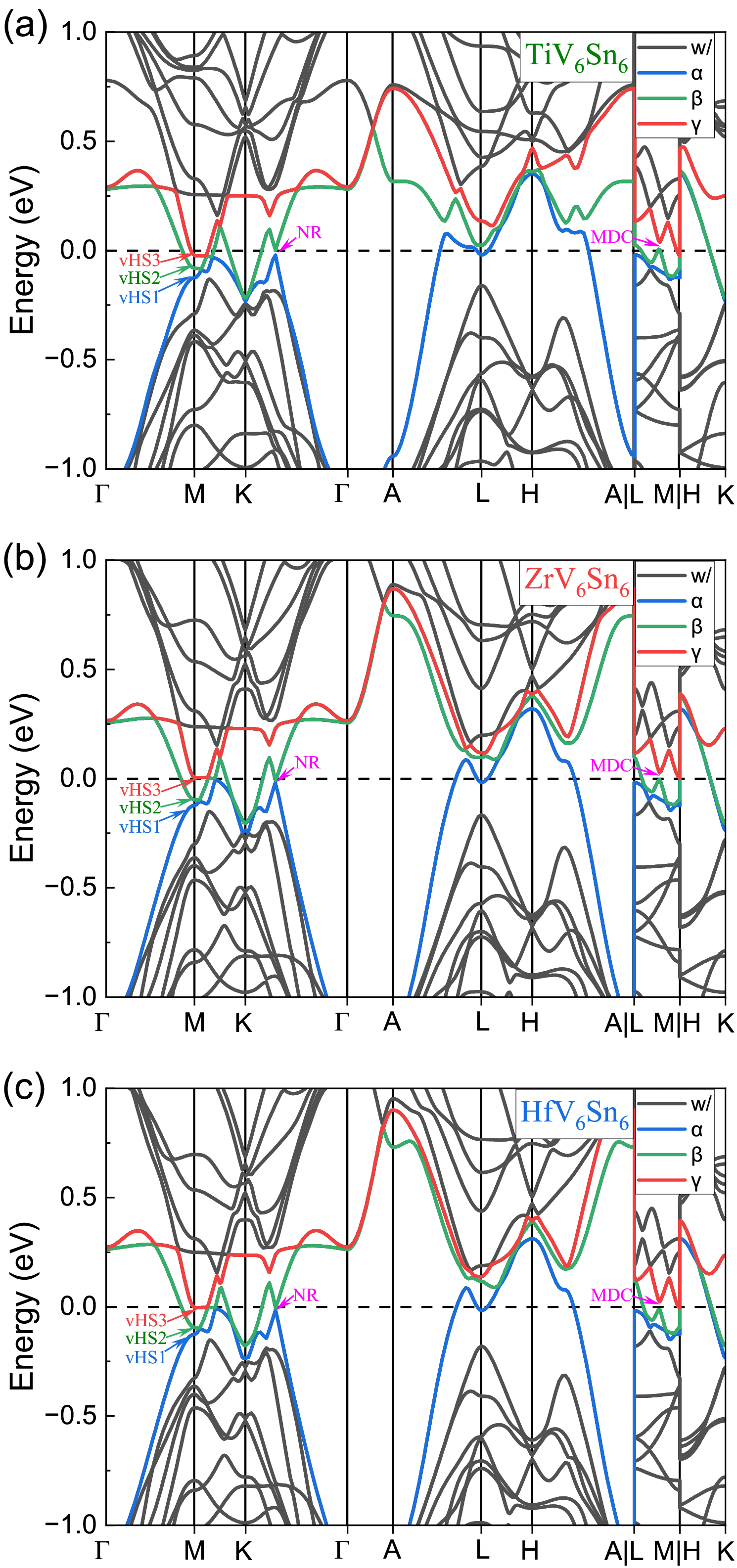}
\caption{
(a)-(c) Electronic band structures with SOC (w/) for (Ti, Zr, Hf)V$_6$Sn$_6$.
According to the relative magnitude of the energy, we labeled the three energy bands near the Fermi level as $\alpha$, $\beta$, and $\gamma$.
Near the Fermi level are highlighted three vHSs at $M$, a nodal ring (NR) intercepted at the $K$-$\Gamma$ line, and a massive Dirac cone (MDC) along the $L$-$M$ line.
}
\label{f4}
\end{figure}

\begin{figure}[h]
\centering
\includegraphics[clip, width=0.5\textwidth]{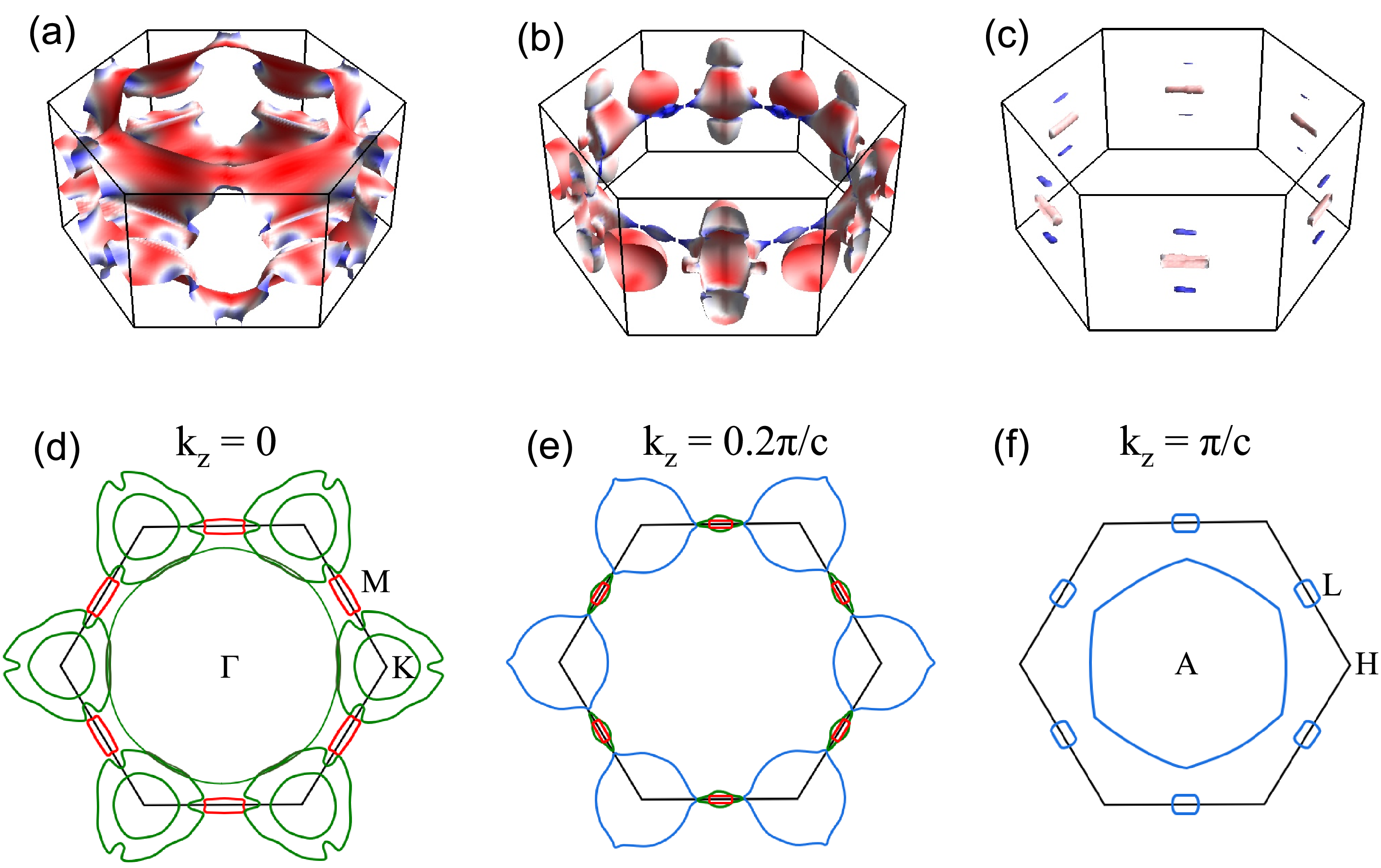}
\caption{
(a)-(c) The Fermi surfaces of ZrV$_6$Sn$_6$ in the first Brillouin zone for three bands crossing the Fermi level.
(d)-(f) The cross sections of Fermi surfaces at $k_z=0$, $k_z=0.2\pi/c$, and $k_z=\pi/c$ slices for ZrV$_6$Sn$_6$.
}
\label{f5}
\end{figure}

The most remarkable feature observed in MR and $\rho_{yx}$ curves is the presence of SdH QOs.
These quantum oscillations are particularly pronounced in ZrV$_6$Sn$_6$, and persist even at temperatures as high as 50~K.
After subtracting a smooth background signal, the oscillatory components as a function of the inverse magnetic field 1/B are shown in Figs.~\ref{f3}(a)-~\ref{f3}(c).
As expected, the oscillations in 1/B exhibit perfect periodicity with the oscillatory phase fixed at different temperatures, which originate from the quantization of Landau energy levels.
Fast Fourier transform (FFT) of the SdH QOs [Fig.~\ref{f3}(d)-~\ref{f3}(f)] reveals a single dominant frequency for all three compounds, with $F_{Ti}$ = 205~T, $F_{Zr}$ = 36~T, and $F_{Hf}$ = 33~T, respectively.
In accordance with the Onsager relation $F = (\phi_0/2\pi^2)A_F$ where $\phi_0$ = 2.068$\times$10$^{15}$~Wb is the magnetic flux quantum and $A_F$ is the area of extremal orbit of the Fermi surface~\cite{shoenberg2009}, the determined $A_F$ is 0.020, 0.003, and 0.003~${\AA}^{-2}$ for Ti, Zr, and HfV$_6$Sn$_6$, respectively.
All the cross-sectional areas of Fermi surfaces are very small, demonstrating the presence of tiny pockets in the Brillouin-zone area.

The amplitude of the SdH oscillations, in general, can be described by the Lifshitz-Kosevich (LK) formula with Berry phase taken into account~\cite{shoenberg2009}:
$$\Delta\sigma_{xx} \varpropto R_TR_DR_S\cos\left[2\pi\left(\frac{F}{B}-\frac{1}{2}+\beta\right)\right],$$
where the three coefficients $R_T$, $R_D$, and $R_S$ are the thermal, Dingle, and spin damping factors due to Landau-level broadening, respectively.
The oscillation components are proportional to the cosine term with phase factor $(-\frac{1}{2}+\beta)$, where $\beta = \frac{\phi_B}{2\pi}$, and $\phi_B$ represents the Berry phase.
We show below that our observed SdH QOs can be well fitted to the above LK formula.

As depicted in Fig.~\ref{f3}(h), the temperature dependence of the SdH QOs is well described by the thermal-damping term in the LK formula, i.e., $R_T = \frac{\alpha m^\ast T/B}{\sinh \alpha m^\ast T/B}$, where $\alpha=2\pi^2 k_B/e\hbar$ and $m^\ast$ is the cyclotron mass.
The best fits yield cyclotron masses $m_{Ti}^*=0.35\ m_e$, $m_{Zr}^*=0.035\ m_e$ and $m_{Hf}^*=0.035\ m_e$.
The near-zero cyclotron masses are comparable to the one in YV$_6$Sn$_6$ within the \textit{R}V$_6$Sn$_6$ family, but much smaller than those reported for ScV$_6$Sn$_6$, \textit{R}Mn$_6$Sn$_6$ (\textit{R} = Gd - Er) and \textit{A}V$_3$Sb$_5$ ~\cite{Ma2021rare,Pokharel2021,shrestha2023electronic}, which most likely derive from the low-energy topological excitations.
Figure~\ref{f3}(g) presents the Landau fan diagram constructed from the SdH QOs.
As $\rho_{xx}\gg\rho_{yx}$ in (Ti, Zr, Hf)V$_6$Sn$_6$ in Figs.~\ref{f2}(a)-~\ref{f2}(f), the minima and maxima of the oscillations in $\sigma_{xx}$ are out of phase with those in the $\rho_{xx}$.
Correspondingly, the oscillatory minima of $\rho_{xx}$ are defined as integral Landau indices here, while the maxima as half indices~\cite{Ando2013,Wang2016}.
The Landau index $N$ and 1/B satisfy the Lifshitz-Onsager quantization rule described by $F/B_N = N+\frac{1}{2}-\beta$.
By linear extrapolation to the $N$ axis, the obtained intercepts for the three compounds are all between $\pm$1/8, pointing to the same nontrivial Berry phase~\cite{Wang2016,Mikitik1999,Mikitik2012,Murakawa2013}.

We also fit the field dependence of the oscillation amplitude normalized by $R_T$ to $R_D$ as shown in the Fig.~\ref{f3}(h) inset.
Here $R_D = \exp(-\frac{\alpha m^\ast T_D}{B})$, and $T_D$ is the Dingle temperature.
The obtained $T_D$ is 19.1, 13.6, and 11.3~K for the Ti/Zr/Hf at 1.8~K, from which we can estimate the quantum relaxation time ($\tau_q = \hbar/2\pi k_BT_D$) and the quantum mobility ($\mu_q = e\tau_q/m^*$).
For Zr and Hf siblings, the quantum mobility $\mu_q$ is estimated to 4433~cm$^2$/Vs and 5335~cm$^2$/Vs, respectively, which are even a bit larger than their carrier mobilities ($\sim$10$^3$~cm$^2$/Vs, Fig. S5~\cite{SM}).
Usually the transport relaxation time is not significantly affected by small angular scattering, while the quantum relaxation time is~\cite{PhysRevB.96.045127}.
The large quantum mobilities in  Zr and Hf siblings highlight strong topology protection within the system.
For TiV$_6$Sn$_6$, however, the quantum mobility is reduced to around 300~cm$^2$/Vs, much smaller than its carrier mobility.
Such a difference may lay in the fact that a slight self-doping of the Ti ions into the V kagome sites (as seen from the chemical composition) significantly increases the quantum transport scattering.
Additional estimated parameters for quantum oscillations of the three compounds are listed in Table~\ref{t2}.

To comprehensively understand the experimental results of SdH oscillations, we conduct the first-principles electronic structures calculations with SOC on the (Ti, Zr, Hf)V$_6$Sn$_6$, respectively, as shown in Figs.~\ref{f4}(a)-~\ref{f4}(c).
The three compounds share similar band structures.
The topological features of the kagome lattice are well reproduced, including multiple Dirac cones centered at $K$ which are 0.2~eV below and 0.5~eV above the Fermi level, a quasi-flat band 0.3~eV above the Fermi level, and three vHSs at the $M$ point (originated from the $\alpha$, $\beta$, and $\gamma$ bands, respectively).
An outstanding feature of this system is that the vHS3 locates very close to the Fermi energy level in the band structure (20~meV for Ti and less than 5~meV for Zr and Hf).
This characteristic is thought to be responsible for the many-body electronic orders in the \textit{A}V$_3$Sb$_5$ family~\cite{Ortiz2020,Jiang2021Unconventional,Kang2022,Rina2022}.
There is also one nodal ring intercepted at the $K$-$\Gamma$ line and a massive Dirac cone along the $L$-$M$ line.
These bands together contribute multiple electron- and hole-like sheets of Fermi surfaces as shown in Figs.~\ref{f5}(a)-~\ref{f5}(c) and Fig. S7~\cite{SM}. Note that despite the similar band structures, the Fermi level in TiV$_6$Sn$_6$ is slightly higher compared with Zr/Hf siblings due to a smaller atomic radius and weaker SOC strength, which leads to the differences among the three compounds in the quantum oscillation frequencies and cyclotron masses in Table~\ref{t2}.

We have plotted the cross-section areas for ZrV$_6$Sn$_6$ at three representative slices ($k_z=0$, $k_z=0.2\pi/c$, and $k_z=\pi/c$), as shown in Figs.~\ref{f5}(d)-~\ref{f5}(f).
As the vHS3, the nodal ring and the massive Dirac cone are all close to the Fermi level, they contribute to multiple small Fermi pockets with nonzero Berry curvatures, close to our observed quantum oscillation parameters~\cite{Li2018}.
In addition, as the orbits are very close to each other in the $k_z=0$ plane, there can also exist small breakdown orbits~\cite{shoenberg2009}.
Even combined with angle-dependent electric transport (as shown in Fig.~S6~\cite{SM}), it is hard at this moment to nail down the exact pockets responsible for the quantum oscillations we observed.
Future high-resolution torque measurement will be helpful for complete mapping of the Fermi surfaces.

Although the band structures of the three compounds closely resemble those of ScV$_6$Sn$_6$, which is the sole member displaying the CDW phase and maintains its integrity throughout the charge-order transition at the vHS~\cite{Arachchige2022}, we observe no signs of the CDW and superconductivity in the (Ti, Zr, Hf)V$_6$Sn$_6$ down to 0.3~K in the He3 experiment (in Fig.~S3~\cite{SM}).
It is possible that the CDW or superconducting phases may be potentially achieved via further tuning the vHS to the Fermi level in subsequent experiments involving high pressure or chemical doping.

\section{Conclusion}
In summary, we have successfully synthesized three V-based kagome metals, TiV$_6$Sn$_6$, ZrV$_6$Sn$_6$, and HfV$_6$Sn$_6$, and systematically studied their structures, magnetic, and electrical transport properties.
The single crystals, similar to other nonmagnetic \textit{R}-based \textit{R}V$_6$Sn$_6$, possess a Pauli paramagnetic ground state.
At low temperatures all three compounds exhibit a large unsaturated magnetoresistance and multiband Hall effect.
Distinct SdH QOs with a unique frequency are characterized by near-zero cyclotron masses and nontrivial Berry phases, which is similar to YV$_6$Sn$_6$ and gives clear transport evidence of the relativistic low-energy excitations.
Notably, our first-principles calculations and quantum oscillation parameters suggest that the vHSs at the $M$ point, together with rich topological band crossings, locate just around the Fermi level.
Despite no signs of the CDW and superconductivity in (Ti, Zr, Hf)V$_6$Sn$_6$ in the present research, there is a possibility that the electronic instabilities could be achieved via further tuning the vHS towards the Fermi level in subsequent high pressure or chemical doping experiments.
Our result extends the kagome family and may provide an opportunity in the pursuit of vHS-related physics in kagome metals.

\section{Acknowledgements}
This work was supported by National Key R\&D Program of China Grant No. 2023YFA1407300, National Natural Science Foundation of China Grant No. U2032213, No. 12104461, No. 12374129, and No. 12304156, and No. 12174394, and the Chinese Academy of Sciences under contracts No. YSBR-084 and No.  JZHKYPT-2021-08.
A portion of this work was supported by the High Magnetic Field Laboratory of Anhui Province.
X.X. acknowledges support from Anhui Provincial Natural Science Foundation Grant No. 2108085QA23 and CAS Key Laboratory of Photovoltaic and Energy Conservation Materials Fund Grant No. PECL2021ZD003.
J.Z. acknowledges support from HFIPS Director\texttt{'}s Fund Grant No. BJPY2023B05.

%


%

\clearpage

\end{document}